\documentclass[preprints,article,accept,oneauthor]{mdpi} 
\firstpage{1} 

\pubvolume{vv}
\issuenum{X}
\articlenumber{Y}
\pubyear{2021}
\copyrightyear{2021}
\history{Received: date; Accepted: date; Published: date}
\pdfoutput=1

\Title{Polynomial Distributions and Transformations}

\Author{Yue Yu$^{1}$ and Pavel Loskot$^{1}$*\orcidA{}}

\AuthorNames{Yue Yu and Pavel Loskot}

\address{$^{1}$ \quad ZJU-UIUC Institute, Haining, China;
  \{yue2.22,pavelloskot\}@intl.zju.edu.cn}

\corres{Correspondence: pavelloskot@intl.zju.edu.cn; Tel.: +86-571-8757-2579}

\newcommand{\eref}[1]{(\ref{#1})} 
\newcommand{\scref}[1]{Section~\ref{#1}} 
\newcommand{\sscref}[1]{Subsection~\ref{#1}} 
\newcommand{\dfref}[1]{Definition~\ref{#1}} 
\newcommand{\lmref}[1]{Lemma~\ref{#1}} 
\newcommand{\thref}[1]{Theorem~\ref{#1}} 
\newcommand{\apref}[1]{Appendix~\ref{#1}} 
\newcommand{\apxitem}[1]{%
  \bigskip\noindent\textbf{\textit{#1}:}} 
  
\newcounter{conjecture}
\newcounter{construction}
\newtheorem{Construction}[construction]{Construction}

\usepackage{psfrag}
\usepackage{amssymb,amsfonts,mathrsfs,mathptm,bm,mathtools}

\newcommand{\R}{\mathcal{R}} 
\newcommand{\Cb}{\mathbb{C}}
\newcommand{\Zs}{\mathbf{0}} 
\newcommand{\conv}{\circledast} 
\newcommand{\const}{\operatorname{const}}

\newcommand{\eee}{\,{\operatorname{e}}} 
\DeclareMathOperator*{\argmin}{argmin} 
\DeclareMathOperator*{\argmax}{argmax} 
\newcommand{\jj}{{\operatorname{j}}} 
\newcommand{\df}{{\,\operatorname{d}}} 

\newcommand{\br}{\begin{array}}
\newcommand{\er}{\end{array}}

\newcommand{\E}[1]{{\operatorname{E}}\!\left[#1\right]} 
\newcommand{\var}[1]{{\operatorname{var}}\!\left[#1\right]} 
\newcommand{\cov}[1]{{\operatorname{cov}}\!\left[#1\right]} 

\newcommand{\norm}[1]{\left\lVert#1\right\rVert}

\newcommand{\dotprod}[1]{\left\langle{#1}\right\rangle}
\renewcommand{\Re}[1]{\operatorname{Re}\!\left(#1\right)}
\renewcommand{\Im}[1]{\operatorname{Im}\!\left(#1\right)}
\newcommand{\Ip}{\tilde{p}}
\newcommand{\Dp}{\dot{p}}
\newcommand{\Intn}[2]{\overset{\ #1}{\tilde{#2}}}
\newcommand{\Dern}[2]{\overset{\ #1}{\dot{#2}}}

\newcommand{\vy}{\bm{y}}
\newcommand{\va}{\bm{a}}
\newcommand{\vb}{\bm{b}}

\newcommand{\vr}{\bm{r}}
\newcommand{\vx}{\bm{x}}
\newcommand{\vX}{\bm{X}}
\newcommand{\vw}{\bm{w}}
\newcommand{\vJ}{\bm{J}}
\newcommand{\cL}{\mathcal{L}}

\newcommand{\KL}{\mathrm{KL}}
\newcommand{\cM}{\mathcal{M}}
\newcommand{\vM}{\bm{M}}
\newcommand{\vB}{\bm{B}}

\abstract{Polynomials are common algebraic structures, which are often used to
  approximate functions including probability distributions. This paper
  proposes to directly define polynomial distributions in order to describe
  stochastic properties of systems rather than to assume polynomials for only
  approximating known or empirically estimated distributions. Polynomial
  distributions offer a great modeling flexibility, and often, also
  mathematical tractability. However, unlike canonical distributions,
  polynomial functions may have non-negative values in the interval of support
  for some parameter values, the number of their parameters is usually much
  larger than for canonical distributions, and the interval of support must be
  finite. In particular, polynomial distributions are defined here assuming
  three forms of polynomial function. The transformation of polynomial
  distributions and fitting a histogram to a polynomial distribution are
  considered. The key properties of polynomial distributions are derived in
  closed-form. A piecewise polynomial distribution construction is devised to
  ensure that it is non-negative over the support interval. Finally, the
  problems of estimating parameters of polynomial distributions and generating
  polynomially distributed samples are also studied.}

\keyword{Approximation; distribution; histogram; least-squares; polynomial;
  probability density}

\begin{document}

\section{Introduction}\label{sc:1}

Approximating functions is motivated by reducing the computational complexity
and achieving analytical tractability of mathematical models. This also
includes the problems of finding the low-complexity and low-dimensional
mathematical models for continuous or discrete-time observations such as
time-series data and empirically determined features such as histograms. This
paper is concerned with the latter problem, i.e., how to effectively model the
probability distributions of observation data. In particular, it is proposed to
define polynomial probability distributions rather than to assume polynomial
approximations of probability distributions. This is a major departure from
reasoning found in the existing literature.

Polynomial distributions provide a superior flexibility over other canonical
distributions, albeit at a cost of larger number of parameters, and the support
interval is constrained to a finite range of values. The main advantage of
polynomial distributions is that they can yield parameterized closed-form
mathematical expressions as well as offer a much greater flexibility in
modeling time-evolution of probability distributions, for example, when
describing causal interactions in complex systems and modeling state
transitions in dynamic systems. This may lead to development of novel
probabilistic mathematical frameworks involving polynomial distributions. The
disadvantage is that, in case of a general polynomial function, it may be
difficult to ensure that the polynomial is non-negative over the whole indented
interval of support. However, the non-negativity can be guaranteed, for
example, by assuming the squared polynomials.

Weierstrass theorem \cite{Phillips1996} is the fundamental result in
approximation theory of functions. It states that every continuous function can
be uniformly approximated with an arbitrary precision over any finite interval
by a polynomial of a sufficient order. The uniform approximation can be
expressed as a sequence of algebraic polynomials uniformly converging over a
given interval to the function of interest. The approximation accuracy can be
evaluated by different metrics including $l_p$-norms, minimax norm and other.
The best approximating function from a set or from a sequence of functions and
its properties can be determined by Jackson theorem. Stone-Weierstrass theorem
generalizes the function approximation to cases of multivariate functions and
functions in multiple dimensions \cite{Cheney1982}.

Polynomials can be used to approximate known probability distributions as well
as distributions estimated as a histogram \cite{Freedman1981, Munkhammar2017}.
The paper \cite{Badinelli1996} is one of the earlier works assuming the
approximation of probability distributions by a polynomial. Fitting of
multivariate polynomials to multivariate cumulative distributions and their
partial derivatives is studied in \cite{Abdous2007}, whereas multivariate
polynomial interpolation is studied in \cite{Gasca2000}. The conditions for the
coefficients of a polynomial to be a sum of two squared polynomials are
determined in \cite{Ghasemi2010}.

The problem of fitting a polynomial into a finite number of data samples has
been investigated in the classic reference \cite{Forsythe1957}. The polynomial
curve fitting methods are often available in various software packages
\cite{Cunis2018}. Modeling times-series data by piecewise polynomials is
considered in \cite{Hiang2013} and in \cite{Gao2020}. The least-squares
polynomial approximation of random data samples with standard and induced
densities is compared in \cite{Guo2020}. A new method for polynomial
interpolation of data points within a plane is proposed in \cite{Han2007}.
Interestingly, the recent survey \cite{Melucci2019} on approximating
probability distributions does not mention polynomial approximation as one of
the available methods.

The polynomial expansion of chaos for reliability analysis of systems is
proposed in \cite{He2021}. A polynomial kernel for feature learning from data
is considered in \cite{Chen2015}. Stone-Weierstrass theorem is assumed in
\cite{Cotter1990} to design a neural network that can approximate an arbitrary
measurable function. The method for function approximation by a polynomial
using a neural network is investigated in \cite{Tong2021}.

Polynomials can be sparse, i.e., only some of their coefficients including the
coefficient determining their order are non-zero. The polynomials with special
properties have been named; for example, there are Lagrange, Legendre, Diskson,
Chebyshev and Bernstein polynomials \cite{Barbeau1989, Rahman2002}. The special
polynomials such as Hermite and Lagrangian polynomials can form a basis for
function decomposition. There is a close link with approximating periodic
continuous functions by trigonometric polynomials in the Fourier analysis
\cite{Apostol1974}. A procedure for orthogonal polynomial decomposition of
multivariate distributions has been devised in \cite{Rahman2009} in order to
compute the output of a multidimensional system with a stochastic input.

The reference \cite{Barbeau1989} is a comprehensive textbook on theory of
polynomials covering fundamental theorems, special polynomials, polynomial
algebra, finding and approximating polynomial roots, finding polynomial
factors, solving polynomial equations, and defining polynomial inequalities and
properties of polynomial approximations. The other textbook \cite{Rahman2002}
includes additional topics such as critical points of polynomials, composition
of polynomial functions, theorems and conjectures about polynomials, and
defining extremal properties of polynomials. Although the textbook
\cite{Funaro1992} focuses on solving differential equations by polynomial
approximations, it also provides a necessary background on polynomials
including their definitions and properties. Differential equations are solved
by Jacobi polynomial approximation in \cite{Guo2009}.

The properties of minima and maxima of polynomials were studied in
\cite{Boas1977}. An algorithm for finding the global minimum of a general
multivariate polynomial has been developed in \cite{Hanzon2003}. The number of
local minima of a multivariate polynomial is bounded in \cite{Qi2003}. Sturm
series are assumed in \cite{Uteshev1998} to find the maxima of a polynomial.

In this paper, polynomial distributions are introduced in \scref{sc:2}
including transformations of polynomial distributions, fitting a histogram with
a polynomial distribution, constructing piecewise polynomial distributions, and
presenting basic properties of random polynomial functions. In \scref{sc:3},
selected properties of polynomial distributions are derived. The estimation
problems involving polynomial distributions are considered in \scref{sc:4}, and
the problem of generating polynomially distributed random variables is
discussed in \scref{sc:5}. The paper is concluded in \scref{sc:6}. Furthermore,
the key expressions for polynomial functions in Form I, II and III are
summarized in Appendix A, B and C, respectively.

The following notations are used in the paper: $X$ denotes a random variable
whereas $x$ denotes a specific values of this random variable; $(\cdot)^T$ is
matrix transpose, $(\cdot)^{-1}$ is matrix inverse, and the operators,
$\E{\cdot}$ and $\var{\cdot}$, denote expectation and variance, respectively.

\section{Defining polynomial distributions}\label{sc:2}

Given a continuous interval, $(l,u)\subseteq \R$, the probability density
function (PDF), $p(X)$, of a random variable, $X$, with the support, $(l,u)$,
must satisfy the following two conditions,
\begin{equation}\label{eq:5}
  \begin{split}
    p(x) & \geq 0, \quad \forall x\in(l,u) \\
    \int_l^u p(x) & \df x = 1.
  \end{split}
\end{equation}
Assume that the PDF, $p(X)$, can be linearly expanded as,
\begin{equation}\label{eq:10}
  p(x) = a_0 + \sum_{i=1}^n  a_i b_i(x)
\end{equation}
into a $n$-dimensional basis of generally non-linear functions, $b_i(x)$.
Provided that the functions, $b_i(x)$, are themselves PDFs, i.e., they satisfy
conditions \eref{eq:5}, and, $\sum_{i=0}^N a_i=1$, the PDF \eref{eq:10} is
referred to as mixture distribution. Alternatively, it is possible to assume
the parameterization, $b_i(x) \equiv b(x;\theta_i)$.

In this paper, let, $b_i(x)=x^i$, so that, the expression \eref{eq:10}
represents an ordinary univariate polynomial of degree, $n$. The coefficients,
$a_i$, can be a function of another common variable, e.g., $a_i(y)$,
$i=0,1,\ldots,n$; such a multivariate polynomial is referred to as algebraic
function. The multivariate polynomial having the same degree of non-zero term
is referred to as being homogeneous (formerly a quantic polynomial).

The following three representations of real-valued polynomial functions are
considered in this paper.
\begin{Definition}\label{df:1}
  \begin{subequations}
    \begin{eqnarray}
      \label{eq:20a}\mathrm{Form\ I:} & p_n(x)=\sum\limits_{i=0}^n a_i\,x^i, &
      a_i\in\R,\ a_n\neq 0\\
      \label{eq:20b}\mathrm{Form\ II:} &  p_n(x) = a_n \prod \limits_{i=1}^n
      (x-r_i), & r_i\in\Cb,\ a_n\neq 0 \\
      \label{eq:20c}\mathrm{Form\ III:} &   p_n(x) = \sum\limits_{i=1}^n
      \frac{a_i}{x-r_i}, & a_i\neq 0,\ r_i\neq r_j\ 
      \forall i\neq j
    \end{eqnarray}
  \end{subequations}  
\end{Definition}

Form I is a canonical polynomial function. Form II indicates that every
$n$-degree polynomial has exactly $n$, generally complex-valued, roots $r_i$
\cite{Pan1997}. The number of real-valued roots can be determined by Sturm's
theorem. Form III is a rational polynomial function. The basic properties of
the polynomial Forms I, II and III are summarized in Appendix A, B and C,
respectively, including roots, indefinite and definite integrals, derivatives,
general statistical moments, and characteristic or moment generating functions.
Note that every polynomial function, $p_n(x)$, of any order, $n$, diverges when
its argument, $x$, becomes unbounded. In addition, Forms I and II are
equivalent as shown in \apref{Apx:B}, and, for complex-conjugate roots,
$(x-r_i)(x-r^\ast_i)= (x-\Re{r_i})^2+ \Im{r_i}^2>0$. Form I defined by
\eref{eq:20a} can be also computed recursively as,
\begin{equation}
  \begin{split}
    p_n(x) &= (\dots(((a_nx+a_{n-1})x+a_{n-2})x+a_{n-3}) \dots )x + a_0 \\
    &= x p_{n-1}(x) + a. \\
  \end{split}
\end{equation}

More importantly, the polynomial forms in \dfref{df:1} represent a PDF,
\emph{if and only if}, they satisfy both conditions \eref{eq:5}. This can be
achieved by using linear and non-linear transformations, which are defined in
the following lemma.

\begin{Lemma}\label{lm:1}
  A Form I and II polynomial, $p_n(x)$, of degree $n$ and all its derivatives,
  $p^{(k)}_n(x)$, $k\leq n$, are continuous and strictly bounded over a finite
  interval, $x\in(l,u)$. Then, for any such polynomial, $p_n(x)$,
  \begin{itemize}
  \item[(a)] There exist finite real constants, $A$ and $B$, such that the
    linearly transformed polynomial, $A p_n(x)+B$, satisfies PDF conditions
    \eref{eq:5}.
  \item[(b)] There exists a real positive constant, $A>0$, such that the
    polynomial, $A|p_n(x)|$, or, $A(p_n(x))_{+}$, satisfies PDF conditions
    \eref{eq:5} where $|\cdot|$ denotes absolute value, and $(\cdot)_{+}$
    changes negative values of its argument to zero.
  \item[(c)] There exists a low-degree polynomial, $q_k(x)$, such that the
    polynomial, $q_k(p_n(x))$, satisfies PDF conditions \eref{eq:5}; for
    instance, $q_2(x)=A x^2$, $A>0$.
  \end{itemize}
\end{Lemma}

The polynomial PDFs defined in \lmref{lm:1} can be further constrained by the
required number of local minima, maxima and roots within the interval of
support, $(l,u)$. There can be also additional constraints on smoothness
expressed in terms of the minimum required polynomial order.

By Bolzano's theorem, a continuous function having opposite sign values in an
interval also has a root between these values. Consequently, a polynomial,
$p_n(x)$, of order $n$ have at least one maximum or minimum between every two
adjacent roots, and there can be a maximum or minimum located at roots
themselves \cite{Boas1977}. Moreover, provided that the polynomial is
considered over a finite interval, the boundary points of the support interval
should be treated as additional roots, i.e., the boundary points can create
local maximum or minimum as well as allow additional extrema to exist before
the first nearest root. In case of Form II polynomials, the condition of the
first derivative to be zero can be equivalently expressed as,
\begin{equation}\label{eq:66}
  \frac{\df}{\df x} p_n(x) \overset{!}{=} 0 \quad \Leftrightarrow \quad
  \frac{\df}{\df x} \log p_n(x) = \frac{\Dp_n(x)}{p_n(x)} =
  \sum_{i=1}^n \frac{1}{x-r_i} \overset{!}{=} 0.
\end{equation}
However, this approach still requires finding the roots of \eref{eq:66} for
every sub-interval, $(r_i,r_{i+1})$, $i=0,1,\ldots,n$, where $r_0\equiv l$ and
$r_{n+1}\equiv u$. It may be much easier to find the local extrema by
considering the recursion,
\begin{equation}
  p_n(x) =  \int p_{n-1}(x) \df x = \sum_{i=0}^{n-1} \frac{a_i}{i+1} x^{i+1} + c 
\end{equation}
provided that the roots of the polynomial,
$p_{n-1}(x) = a_{n-1}\prod_{i=1}^{n-1}(x-r_i)$, are known, and, $c=a_0$,
denotes the constant of integration. These roots can be known by design, i.e.,
the locations of minima and maxima are selected a priori in a given interval of
support. More importantly, in case of polynomial PDF, the local maxima
represent the modes of such a distribution.

The Form I polynomial PDF can be generalized as,
\begin{equation}\label{eq:70}
  p_n(x) = \sum_{i=0}^n a_i g^i(x),\quad x\in(l,u)
\end{equation}
where $g(x)$ is a mathematical expression (i.e., not a transformation). For
instance, it is possible to assume polynomials with fractional rather than
integer powers of the independent variable \cite{Beji2021}.

For $g(x)= \eee^{\jj \omega_0 x}$, $\jj=\sqrt{-1}$, $\omega_0=2\pi/(u-l)>0$,
the PDF \eref{eq:70} becomes the truncated exponential Fourier series, i.e.,
\begin{equation}
  p_n(x) = \sum_{i=0}^n a_i \eee^{\jj \omega_0 i x},\quad
  a_i = \frac{1}{u-l} \int_l^u p_n(x) \eee^{-\jj\omega_0 i x}\df x.
\end{equation}
The corresponding $k$-th general moments are then computed as,
\begin{equation}
  \int_l^u x^k\,p_n(x)\df x = \sum_{i=0}^n a_i \int_l^u x^k \eee^{\jj \omega_0
    i x} \df x = \sum_{i=0}^n a_i (-1)^k W^{(k)}(\jj \omega_0 i)
\end{equation}
where $W(\jj\omega)= \int_l^u \eee^{\jj\omega x}\df x$ is the Fourier transform
of a rectangular window located over the interval, $(l,u)$.

For $g(x)= \eee^{x}$, PDF \eref{eq:70} becomes,
\begin{equation}
  p_n(x) = \sum_{i=0}^n a_i \eee^{i x},\quad x\in(l,u)
\end{equation}
which can be readily integrated, although general statistical moments can only
be expressed as special functions.

Consider now the general case of PDF \eref{eq:70}, for $n=2$. Thus, given
$p_2(x)$ and $g(x)$, and positive integers $i_1$ and $i_2$, the task is to find
the coefficients $a_0$, $a_1$ and $a_2$ of the PDF decomposition,
\begin{equation}\label{eq:80}
  p_2(x) = a_2 g^{i_2}(x) + a_1 g^{i_1}(x) + a_0, \quad x\in(l,u).
\end{equation}
Multiplying both sides of \eref{eq:80} by $g^{-i}(x) \dot{g}(x)$ and
integrating, we obtain,
\begin{equation}\label{eq:85}
  \int_l^u p_2(x) g^{-i}(x) \dot{g}(x) \df x = a_2 \int_l^u g^{i_2}(x)
  g^{-i}(x) \dot{g}(x) \df x + a_1 \int_l^u g^{i_1}(x) g^{-i}(x) \dot{g}(x)
  \df x + a_0 \int_l^{g(u)} g^{-i}(x) \dot{g}(x) \df x.
\end{equation}
Assuming a substitution, $y = g(x)$, eq. \eref{eq:85} can be rewritten as,
\begin{equation}
  \int_{g(l)}^{g(u)} p_2(g^{-1}(y)) y^{-i} \df y = a_2
  \int_{g(l)}^{g(u)} y^{i_2-i} \df y + a_1 \int_{g(l)}^{g(u)} y^{i_1-i} \df
  y + a_0 \int_{g(l)}^{g(u)} y^{-i} \df y. 
\end{equation}
Provided that, $g(u)=-g(l)=v$, and $i_1>0$ is an odd-integer, and $i_2>0$ is an
even-integer, then, for $i=i_1$ and $i=i_2$, respectively,
\begin{equation}
  \begin{split}
     \int_{-v}^v p_2(g^{-1}(y)) y^{-i_1} \df y &= a_2
    \underbrace{\int_{-v}^v y^{i_2-i_1} \df y}_{=0} + 2 a_1 v + a_0
    \underbrace{\int_{-v}^v y^{-i_1} \df y}_{=0} \\
    \int_{-v}^v p_2(g^{-1}(y)) y^{-i_2} \df y &= 2 a_2 v + 
    a_1 \underbrace{\int_{-v}^v y^{i_1-i_2} \df y}_{=0} + a_0\int_{-v}^v
    y^{-i_2} \df y
  \end{split}
\end{equation}
and, therefore,
\begin{equation}\label{eq:90}
  \begin{split}
    a_1 &= \frac{1}{2v} \int\limits_{-v}^v p_2(g^{-1}(y)) y^{-i_1} \df y \\
    a_2 &= \frac{1}{2v} \int\limits_{-v}^v \left(p_2(g^{-1}(y))-a_0\right)
    y^{-i_2} \df y.    
  \end{split}
\end{equation}
The offset, $a_0$, must be computed from some other constraint, for example, as
the minimum value to guarantee a non-negativity of $p_2(x)$. Note that the
function, $g(x)$, in \eref{eq:80} must be chosen, so the integrals \eref{eq:90}
converge.

\subsection{Probability density transformations}\label{ssc:10}

In general, if $g(X)$ is an invertible memoryless transformation of a random
variable, $X$, having the PDF, $p(X)$, the PDF, $q(X)$, of random output
variable, $g(X)$, is, \cite{Papoulis02}
\begin{equation}\label{eq:60}
  q(x) = \frac{p(g^{-1}(x))}{|\dot{g}(g^{-1}(x))} =
  p(g^{-1}(x)) \left| \frac{\df g^{-1}(x)}{\df x} \right|.
\end{equation}
Assuming $p(x)$ is a Form I polynomial PDF, $p_n(x)$, the transformed PDF is
also a Form I polynomial, i.e.,
\begin{equation}\label{eq:60a}
  q_n(x) =  |\dot{g}^{-1}(x)|\, \sum_{i=1}^n a_i g^{-i}(x)
  = \sum_{i=1}^n a_i \left( \frac{\sqrt[i]{|\dot{g}^{-1}(x)|}}{g(x)} \right)^i
\end{equation}
in variable, $\sqrt[i]{|\dot{g}^{-1}(x)|} g^{-1}(x)$. 

Assuming a linear transformation, $g_1(x)=b_1x+b_0$, the PDF \eref{eq:60a} is
also a polynomial PDF of the same order, i.e.,
\begin{equation}
  q_n(x) = \sum_{i=1}^n \frac{a_i}{|b_1|b_1^i} (x-b_0)^i.
\end{equation}
However, the linear transformation changes the support, $(l,u)$ of $p_n(x)$, to
$(b_1l+b_0,b_1u+b_0)$, if $b_1>0$, and $(b_1u+b_0,b_1l+b_0)$, if $b_1<0$.

Another example of a non-linear transformation with memory that preserves
polynomial form of the resulting distribution is an integrator. In particular,
let, $g^{-1}(x) = \int_{-\infty}^x f(u)\df u \equiv F(x)$, i.e.,
$g(x)=F^{-1}(x)$, such that, $f(u)\geq 0$, for $\forall u$. Then, substituting
into \eref{eq:60}, the transformed PDF can be written as,
\begin{equation}\label{eq:150}
  q_m(x) = b_1 f(x) p_n\left( b_1 F(x) + b_0 \right) 
\end{equation}
where $b_1\neq 0$ and $b_0$ are arbitrary real constants. Provided that $f(x)$
is a polynomial of order, $k$, $F(x)$ is a polynomial of order, $(k+1)$ (cf.
\apref{Apx:A}), and thus, $m= n(k+1)k$. The family of PDFs with a form similar
to \eref{eq:150} have been considered in \cite{Alzaatreh2013}, which could be
investigated also for our case of polynomial distributions.

Consider now a general case of a polynomial nonlinear transformation, $g_k(x)$,
and denote as $x_i$, $i=1,2,\ldots,N(y)$, all the roots of, $g_k(x)=y$. Then,
the PDF \eref{eq:60} is rewritten as,
\begin{equation}
  q_m(y) = \sum_{j=1}^{N(y)} \frac{p_n(x_j(y))}{|\dot{g_k}(x_j(y))|}
\end{equation}
i.e., it is a sum of ratios of polynomials, i.e., $q_m(y)$ is a polynomial of a
certain order, $m$.

Linear and non-linear transformations of a random variable can be used to
change the interval of support of its probability distribution. The following
\lmref{lm:2} assumes linear transformations to convert the interval of support,
$(l,u)$, into $(-1,+1)$ and vice versa. \lmref{lm:3} proposes two
transformations how to convert the interval of support, $(-1,+1)$, to
semi-finite or infinite intervals of support, respectively.

\begin{Lemma}\label{lm:2}
  The interval of support, $(l,u)$, of a PDF, $p(x)$, is changed to the
  interval, $(-1,+1)$, by a linear transformation,
  $\frac{2}{u-l}X - \frac{u+l}{u-l}$, which transforms the PDF, $p(x)$, to the
  PDF, $\frac{u-l}{2} p\left( \frac{(u-l)x-(u+l)}{2}\right)$. Furthermore, the
  linear transformation, $\frac{(u-l)}{2}X+\frac{u+l}{2}$, transforms the PDF,
  $p(x)$, with support, $(-1,+1)$, into the PDF,
  $\frac{2}{u-l} p\left( \frac{2X-(u+l)}{u-l} \right)$, with the interval of
  support, $(l,u)$.
\end{Lemma}

\begin{Lemma}\label{lm:3}
  The PDF, $p(x)$, defined over the finite interval of support, $(-1,+1)$, can
  be transformed into the PDF,
  $(x^2+x+1/4)^{-1}p\left(\frac{2x-1}{2x+1}\right)$, with a semi-infinite
  support, $(0,+\infty)$, using the non-linear transformation,
  $\frac{1}{2} \left( \frac{1+X}{1-X}\right)$. Similarly, the non-linear
  transformation, $\mathrm{atanh}(X)$, can be assumed to extend the support to
  all real numbers, for a PDF, $p(x)$, defined over the support interval,
  $(-1,+1)$. The transformed PDF becomes,
  $\mathrm{cosh}^{-2}(x) p\left( \mathrm{tanh}(x)\right)$.
\end{Lemma}

\subsection{Polynomial PDF fit of a histogram}

Approximation of a continuous function by a polynomial over a finite interval
is formalized by well-known Weierstrass theorem \cite{Phillips1996}. The
polynomial approximation represents the problems of existence as well as
uniqueness of such a polynomial, and also the problem how to find it. These
problems are crucially dependent on the choice of metric for goodness of
approximation. Hence, consider the problem of approximating a PDF having a
finite support by a polynomial PDF. For instance, empirical histogram can be
fitted by a polynomial function, or a known PDF can be approximated by a
polynomial in order to achieve mathematical tractability. However, in neither
of these cases, the resulting polynomial is guaranteed to satisfy conditions
\eref{eq:5}, since the polynomial coefficients are normally chosen to obtain
the best fit.

The polynomial PDF can be obtained by assuming a polynomial function, which is
non-negative over a given interval for any values of its coefficients. One
example of such a polynomial is, $p_n^2(x)$, which has degree, $2n$. The true
PDF, $q(x)$, can be then approximated as,
\begin{equation}\label{eq:95}
  q(x)\approx p_n^2(x),\quad \mbox{or}, \quad \sqrt{q(x)} \approx p_n(x).
\end{equation}
The latter strategy by first transforming $q(x)$ with a square-root is
numerically more stable. Other such invertible transformations of $q(x)$ can
also be assumed, provided that they yield non-negative polynomial, $p_n(x)$,
since scaling $p_n(x)$ to have a unit area usually does not affect the
approximation error significantly.

For instance, the data points, $(x_i,\sqrt{y_i})$, $i=1,2,\ldots,M$, can be
interpolated by Lagrange polynomials,
\begin{equation}
  L_i(x)=\prod_{j=1 \atop i\neq j}^M \frac{x-x_i}{x_i-x_j}.
\end{equation}
Then, the true PDF, $p(x)$, is approximated as,
\begin{equation}\label{eq:100}
  p(x) \approx \left( \sum_{i=1}^M\, \sqrt{y_i} \, L_i (x) \right)^2 \equiv
  q_{2(M-1)}(x)
\end{equation}
which is a polynomial of order, $2(M-1)$. In order to normalize the
approximation \eref{eq:100}, let,
\begin{equation}
  c_{ij} = \prod\limits_{j_1=1 \atop j_1\neq i}^M
      \prod\limits_{j_2=1 \atop j_2\neq j}^M (x_i-x_{j_1})(x_j-x_{j_2})
\end{equation}
and, 
\begin{equation}
  \begin{split}
    s_{ij}=\int_l^u L_i(x)L_j(x) \df x &= c_{ij}^{-1} \int_l^u \prod_{j_1=1
      \atop j_1\neq i}^M \prod_{j_2=1 \atop j_2\neq j}^M (x-x_{j_1})
    (x-x_{j_2}) \df x\\ &= c_{ij}^{-1}  \sum_{k=0}^{2(M-1)} a_k \int_l^u x^k
    \df x = c_{ij}^{-1}  \sum_{k=0}^{2(M-1)} \frac{a_k}{k+1} (u^{k+1}-l^{k+1}).
  \end{split}
\end{equation}
Then, the area,
\begin{equation}
  \int_l^u q_{2(M-1)}(x)\df x= \sum_{i=1}^M\sum_{j=1}^M \sqrt{y_iy_j}\, s_{ij}.
\end{equation}

The most common method for fitting a polynomial to a histogram is linear
regression \cite{Maulud2020}. Denote the vectors, $\vy=[y_i]$,
$i=1,2,\ldots,M$, and, $\va=[a_j]$, $j=0,1,\ldots,n$, and the matrix,
$\vX=[x_i^j]$. The constrained least squares (LS) problem is then formulated
as,
\begin{equation}\label{eq:110}
  \min_{\va} \norm{\vy - \vX\va}^2,\quad \mbox{s.t.}\quad \vw^T \va=1
\end{equation}
where the weights, $ w_i = \frac{1}{i+1}(u^{i+1}-l^{i+1})$, assuming the
support interval, $(l,u)$. The first derivative of the corresponding Lagrangian
is set equal to zero, and the estimated coefficients, $\hat{\va}$, of the
fitting polynomial, $p_n(x)$, are computed as,
\begin{equation}\label{eq:140}
  \begin{split}
    \frac{\df}{\df \va} \cL(\lambda) &= 2 \vX^T \vX\va - 2 \vX^T\vy +
    \lambda \vw ^T \overset{!}{=} \Zs \\
    \Rightarrow\quad \hat{\va} &= (\vX^T\vX)^{-1} \left(\vX^T\vy +
      \frac{\lambda}{2} \vw^T \right).
  \end{split}
\end{equation}

In order to approximate a known continuous distribution, $f(x)$, over a finite
interval, $(l,u)$, representing the full or truncated support of that
distribution, the constrained least-squares \eref{eq:140} can be again used
assuming the distribution samples, $f(l+\Delta_xi)$, $i=0,1,\ldots$.

If $p_n(x)$ is the best polynomial fit of a histogram, or of a sampled known
PDF, then it must be evaluated whether it is non-negative over the whole
support of interest, $(l,u)$. This can be readily and reliably tested by
numerically computing the integral, $I_1= \Im{\int_l^u \sqrt{p_n(x)} \df x}$,
or, $I_2= \int_l^u p_n(x) - |p_n(x)|\df x$. If $p_n(x)$ contains negative
values within the interval, $(l,u)$, then $I_1\neq 0$, and $I_2<0$,
respectively. It is also possible to assume logarithm instead of square-root in
the definition of integral, $I_1$.

In case the polynomial fitted to a histogram contains negative values, a
constant, $d>0$, can be added to the observed data points, i.e.,
$y_i \mapsto \frac{y_i+d/\Delta_x}{1+Md}$, where $\Delta_x= x_{i+1}-x_i$, and
the scaling ensures that, $\Delta_x \sum_{i=1}^M y_i=1$. Correspondingly, the
fitted polynomial is also shifted and scaled as,
$p_n(x) \mapsto \frac{p_n(x)+d/\Delta_x}{1+dM}$, so $\int_l^u p_n(x)\df x =1$.

Furthermore, the roots of a polynomial, $p_n(x)$, can be constrained in order
to guarantee that it is non-negative over a finite interval, $(l,u)$. This is
formulated in the following theorem.

\begin{Theorem}\label{th:1}
  A Form II real-valued polynomial, $p_n(x)$, of order $n$ with $a_n>0$ and the
  roots, $r_1\leq r_2\leq \ldots \leq r_n$, is non-negative over the interval,
  $(l,u)$, provided that all its roots satisfy at least one of the following
  conditions:
  \begin{itemize}
  \item[(a)] a root has even-multiplicity;
  \item[(b)] a root has a complex conjugate pair;  
  \item[(c)] a (real-valued) root is smaller than $l$;
  \item[(d)] a real-valued root has odd-multiplicity and is larger than $u$;
    the number of such roots must be even.
  \end{itemize}
\end{Theorem}

\begin{proof}
  Form II polynomial is a product of linear functions, $(x-r_i)$. Cases (a),
  (b) and (c) are trivial. Case (d) is a combinatorial problem. The roots with
  odd-multiplicity cannot be smaller than $u$. Even if these roots are all
  larger than $u$, then their number must be even in order for their negative
  parts to cancel out for all values smaller than $u$.
\end{proof}

\begin{Corollary}\label{cl:1}
  A Form II real-valued polynomial, $p_n(x)$, of order $n$ with $a_n>0$ and the
  roots, $r_1\leq r_2\leq \ldots \leq r_n$, has negative values in the
  interval, $(l,u)$, provided that there is an odd-number of real-valued roots
  with odd-multiplicity that are greater than $l$, or, there is an even number
  of real-valued roots with odd-multiplicity and at least one such root is
  located between $l$ and $u$.
\end{Corollary}

\thref{th:1} can be also used for Form I polynomials, provided that they are
converted to Form II as indicated in \apref{Apx:B}. Even though the roots
cannot be obtained analytically for polynomials of order $n>4$ (Abel–Ruffini's
theorem), it may be sometimes possible to consider a product,
$\prod_{j} p_{n_j}(x)$, of polynomials of orders, $n_j\leq 4$, for $\forall j$.

\subsection{Piecewise polynomial PDF}

In some applications, a piecewise polynomial curve fitting can be assumed. In
particular, the following construction is proposed to fit a set of $(M+1)$
points, $(x_i,y_i)$, $i=1,\ldots,(M+1)$, $x_i<x_{i+1}$, and $y_i\geq 0$,
representing either a histogram, or samples of a known PDF. The construction
yields a piecewise polynomial PDF, $p_n(x)$, of the same order $n$, over the
interval, $(l,u)$, with $l=x_1$ and $u=x_{M+1}$, such that, exactly,
$p_n(x_i)=y_i$. The data points, $(x_i,y_i)$, are referred to as control points
of the piecewise polynomial $p_n$.

\begin{Construction} Let $p_n(x)$ be piecewise continuous, and composed of $M$
  non-overlapping polynomial segments, $q_{(i)n}(x)$, i.e.,
  \begin{equation}
    p_n(x) = \sum_{i=1}^{M} w_i\, q_{(i)n}(x).
  \end{equation}
  The segments, $q_{(i)n}(x)$, are strictly increasing, i.e.,
  $q_{(i)n}^\prime(x)>0$, over their support intervals, $(x_i,x_{i+1})$. The
  points, $x_i$, define the local minima and maxima, such that, if $y_i$ is a
  local minimum, then $y_{i+1}$ is a local maximum and vice versa. Then, the
  weights, $w_i=+1$, if $y_i$ is a local minimum, and $w_i=-1$, if $y_i$ is a
  local maximum. In addition, a continuity (smoothness) of order $C$ requires
  that the first $k$ derivatives,
  \begin{equation}
    \lim_{\epsilon\to 0^+} p_n^{(k)}(x+\epsilon) =
    \lim_{\epsilon\to 0^+} p_n^{(k)}(x-\epsilon), \quad \forall
    x\in(x_1,x_{M+1}),\mbox{ and},\ \forall k=0,1,\ldots,C
  \end{equation}
  which needs to be also true at all points of the local minima and maxima,
  i.e., 
  \begin{equation}
    \lim_{\epsilon\to 0^+} q_{(i)n}^{(k)}(x_{i+1}-\epsilon) =
    \lim_{\epsilon\to 0^+} q_{(i+1)n}^{(k)}(x_{i+1}+\epsilon),\quad
    i=1,\ldots,M-1.
  \end{equation}
\end{Construction}

In order to construct the segment polynomials, $q_{(i)n}(x)$, consider two
strictly increasing polynomials, $u_n(x)=\sum_{i=0}^n a_i x^i$, and,
$v_n(x)=\sum_{i=0}^m b_i x^i$, such that, for some $x_0$, the derivatives,
\begin{equation}
  \begin{split}
    u_n^{(k)}(x_0) &= - v_n^{(k)}(x_0),\quad k=0,1,\ldots,C \\
    \sum_{i=k}^n a_i x_0^{i-k} &= -\sum_{i=k}^m b_i x_0^{i-k}    
  \end{split}
\end{equation}
or, in matrix notation,
\begin{equation}\label{eq:30}
  \underbrace{\left[ \begin{array}{ccccc}
      x_0^n & x_0^{n-1} & \cdots & x_0 & 1 \\ x_0^{n-1} & x_0^{n-2} & \cdots &
      1 & 0 \\ & \ddots & & \ddots & \\ x_0^{n-C} & \cdots & 1 & 0 & 0 \\
    \end{array}\right]}_{\vX_{C}(x_0)} \cdot
\underbrace{\left[ \begin{array}{c} a_n \\ a_{n-1} \\ \vdots \\ a_0 \end{array}
  \right]}_{\va} = - \left[ \begin{array}{ccccc}
      x_0^m & x_0^{m-1} & \cdots & x_0 & 1 \\ x_0^{m-1} & x_0^{m-2} & \cdots &
      1 & 0 \\ & \ddots & & \ddots & \\ x_0^{m-C} & \cdots & 1 & 0 & 0 \\
    \end{array}\right]\cdot \underbrace{\left[ \begin{array}{c} b_m \\ b_{m-1}
        \\ \vdots \\ b_0 \end{array} \right]}_{\vb}.
\end{equation}
For $m=n$, eq. \eref{eq:30} can be rewritten as,
\begin{equation}
  \vX_{C}(x_0)(\va+\vb) = \Zs
\end{equation}
so the coefficients $\va$ and $\vb$ are in the null-space of $\vX_C(x_0)$.

Provided that $\va_{(i)}$ denotes the coefficients of
$q_{(i)n}(x) = \sum_{i=0}^n a_i x^i$, it is required that,
\begin{equation}\label{eq:40}
  \begin{split}
    \vX_{C}(x_2)\,(\va_{(1)} + \va_{(2)}) &= \Zs \\
    \vX_{C}(x_3)\,(\va_{(2)} + \va_{(3)}) &= \Zs \\
    \vdots \qquad\qquad & \\
    \vX_{C}(x_{M})\,(\va_{(M-1)} + \va_{(M)}) &= \Zs. \\
  \end{split}
\end{equation}
Note that the matrices, $\vX_{C}(x_i)$, are computed assuming the control
points, $x_i$.

Given the first vector of coefficients, $\va_{(1)}$, the other coefficient
vectors, $\va_{(i)}$, $i=2,3,\ldots,M-1$, can be computed iteratively using the
underdetermined sets of equations \eref{eq:40}. The numerical feasibility of
this problem requires that the order, $n\gg C$.

The vector, $\va_{(1)}$, must be selected, so that $q_{(1)n}(x_1)=y_1$, and,
$q_{(1)n}(x_{M+1})=y_{M+1}$, and $q_{(1)n}(x)>0$ is $C$-continuous for
$x\in(x_1,x_2)$. Let sample $q_{(1)n}(x)$ at $K$ equidistant points between
$x_1$ and $x_2$. The coefficients $\va_{(1)}$ are then the solution of the
quadratic program,
\begin{equation}\label{eq:50a}
  \begin{array}{c}
    \min \dotprod{\va_{(1)},\va_{(1)}} \\ s.t.\quad \dotprod{\vX_{0}(x_1),
      \va_{(1)}} = y_1,\quad \dotprod{\vX_{0}(x_2), \va_{(1)}} = y_2 \\
    \dotprod{w_1 \vX_{0}(x_1+(k-1) \Delta_1),\va_{(1)}} > 0,\quad
    k=1,2,\ldots,K
  \end{array}
\end{equation}
where $\vX_{0}(x)=[x^n,x^{n-1},\ldots,x,1]$, $\Delta_1 = (x_2-x_1)/(K-1)$ is
the sampling step, and $\dotprod{\cdot,\cdot}$ denotes the dot-product of two
vectors. The other coefficients, $\va_{(i)}$, $i>1$, are computed similarly to
the program \eref{eq:50a}, but with an additional constraint due to
\eref{eq:40}. The extended quadratic program to compute these coefficients is
defined as,
\begin{equation}\label{eq:50b}
  \begin{array}{c}
    \min \dotprod{\va_{(i)},\va_{(i)}} \\ s.t.\quad
    \dotprod{\vX_{0}(x_i),\va_{(i)}} = y_i,\quad
    \dotprod{\vX_{0}(x_{i+1}),\va_{(i)}} = y_{i+1} \\ 
    \dotprod{w_i \vX_{0}(x_i+(k-1)\Delta_i),\va_{(i)}} > 0,\quad
    k=1,2,\ldots,K \\ \vX_{C}(x_i) (\va_{(i-1)}+\va_{(i)}) = \Zs    
  \end{array}
\end{equation}
where $\Delta_i = (x_{i+1}-x_i)/(K-1)$, and $i=2,3,\ldots,M$.

More importantly, quadratic programs \eref{eq:50a} and \eref{eq:50b} require
that the constraints are sufficiently underdetermined, i.e., $n\gg C$,
otherwise the solution may be difficult to find, or even may not exist.
Moreover, the solution is less numerically stable for a linear program than for
a quadratic program, and therefore, the quadratic programs should be
considered.

\subsection{Random polynomial functions}

\sscref{ssc:10} considered a non-linear polynomial transformation, $g_n(x)$,
parameterized by the polynomial coefficients, $a_i$, of a random variable, $X$.
Here, a random polynomial function, $p_n(x)$, in the deterministic variable,
$x\in(l,u)$, are parameterized by random coefficients, $a_i$, i.e.,
\begin{equation}
  y = p_n(x) = \sum_{i=0}^n a_i x^i
\end{equation}
is a random variable. Provided that $a_i$ are independent and distributed as,
$f_{a_i}(a_i)$, the PDF of random variable, $Y$, is given by a multi-fold
convolution,
\begin{equation}
  f_y(y) = \left( \prod_{i=0}^n |x^{-i}| \right) f_{a_0}(y) \conv f_{a_1}(y/x)
  \conv \cdots \conv f_{a_n}(y/x^n)
\end{equation}
since, $a_i x^i$, is distributed as, $|x^{-i}| f_{a_i}(y/x^i)$. Furthermore,
the mean and variance of $Y$, respectively, are,
\begin{equation}
  \E{Y}= \sum_{i=0}^n \E{a_i} x^i,\quad \mbox{and},\quad \var{Y}= \sum_{i=0}^n
  \var{a_i} x^{2i}.
\end{equation}

For instance, if $\E{a_i}=\const$, then, for any $n$ and $x\in(-1,+1)$,
$\sum_{i=0}^n x^i\geq 0$. Similarly, if $\var{a_i}=\const$, then, for any $n$
and $x\in(l,u)$, $l<0<u$, $\sum_{i=0}^n x^{2i}\geq 0$ is a strictly convex
function with the minimum at $x=0$.

The bounds for the number of real roots of random but sparse polynomials were
provided in \cite{Gorav2020}. A numerical method for efficiently finding the
zeros of complex valued polynomials of very large orders has been developed in
\cite{Bini1996}. Another method for a rapid root finding of polynomials is
presented in \cite{Lang1994}.

\section{Derived characteristics of a polynomial distribution}\label{sc:3}

The cumulative distribution function (CDF) can be readily obtained for Form I
polynomial PDF as shown in \apref{Apx:A}, i.e.,
\begin{equation}
  \begin{split}
    P_n(x) &= \int_{l}^x p_n(v)\df v = \sum_{i=0}^n \frac{a_i}{i+1}
    (x^{i+1}-l^{i+1}),\quad x\in(l,u) \\
    &= x \sum_{i=0}^n \frac{a_i}{i+1} x^i - l \sum_{i=0}^n \frac{a_i}{i+1} l^i
    = x \tilde{p}_n(x) - l \tilde{p}_n(l).
  \end{split}
\end{equation}
Note also that, for symmetric support interval, when $u=-l$,
$P_n(u)= \sum_{i=0}^n \frac{a_i}{i+1} (u^{i+1} - (-u)^{i+1})$, so the
normalization of PDF to unity is only affected by even-index coefficients,
$a_i$.

In case of Form II polynomial PDF, it is best to convert it to Form I first as
shown in \apref{Apx:B}.

The median ($q=1/2$), and more generally, the quantile, $X_q$, of a polynomial
distribution is defined as,
\begin{equation}
  P_n(X_q) = X_q \tilde{p}_n(X_q) - l \tilde{p}_n(l) 
  \overset{!}{=} q,\quad 0<q<1.
\end{equation}
Denoting, $P_0(q)= l \tilde{p}_n(l)+q$, the quantile is the unique root of the
polynomial, $x \tilde{p}_n(x) - P_0(q) = a(q) (x - X_q)$.

The expressions for general moments and characteristic or moment generating
functions are derived in \apref{Apx:A}, \ref{Apx:B} and \ref{Apx:C},
respectively.

The Kullback-Leibler (KL) divergence or relative entropy between two polynomial
distributions, $p_n(x)= \sum_{i=0}^n a_i x^i = a_n \prod_{i=1}^n (x-r_i)$, and,
$q_m(x)= b_m \prod_{j=1}^m (x-s_j)$, is defined as,
\begin{equation}
  \begin{split}
    \KL(p_n\|q_m) &= \int_l^u p_n(x) \log \frac{p_n(x)}{q_m(x)} \df x
    = \int_l^u \sum_{i=0}^n a_i x^i \log 
    \frac{a_n\prod_{l=1}^n (x-r_l)}{ b_m \prod_{j=1}^m (x-s_j)} \df x \\
    &= \sum_{i=0}^n a_i \int_l^u x^i \left( \log \frac{a_n}{b_m} + \sum_{l=1}^n
      \log (x-r_l) - \sum_{j=1}^m \log (x-s_j) \right) \df x \\
    &= \sum_{i=0}^n a_i \left(\frac{u^{i+1}-l^{i+1}}{i+1}\log \frac{a_n}{b_m}
      + \sum_{l=1}^n \int_l^u x^i \log(x-r_l)\df x - \sum_{j=1}^m \int_l^u x^i
      \log(x-s_j) \df x \right).
  \end{split}
\end{equation}
The inner integral, $ \int_l^u x^i \log(x-r_l)\df x$, can be expressed in terms
of the hypergeometric, ${}_{2}F_1$, functions with the help of, for example,
\textsf{Mathematica} software.

Differential entropy of a polynomial distribution, $p_n(x)$, is defined as,
\begin{equation}
  \begin{split}
    H(p_n) & = - \int_l^u p_n(x) \log p_n(x) \df x =
    - \int_l^u \sum_{i=0}^n a_i x^i \log a_n \prod_{i=1}^n (x-r_i) \df x \\
    &= \sum_{i=0}^n a_i \left( \frac{\log a_n}{i+1} (u^{i+1}-l^{i+1}) +
      \sum_{l=1}^n \int_l^u x^i \log(x-r_l) \df x \right).
  \end{split}
\end{equation}

Finally, the sum, $Z=X+Y$, of two independent random variables, $X$, and, $Y$,
having the polynomial distributions, $p_n(x)$, and, $q_m(y)$, respectively,
with the same interval of support, $(l,u)$, has also the polynomial
distribution, $f_m(z)$, given by the convolution,
\begin{equation}
  \begin{split}
    f_m(z) &= \int_l^u p_n(x) q_m(z-x) \df x = \int_l^u \sum_{i=0}^n a_i x^i
    \sum_{j=0}^m b_j(z-x)^j \df x \\ &= \sum_{i=0}^n \sum_{j=0}^m a_ib_j
    \int_l^u x^i (z-x)^j \df x = \sum_{i=0}^n \sum_{j=0}^m a_ib_j \sum_{k=0}^j
    \binom{j}{k} z^{j-k} \int_l^u x^{i+k} \df x \\ &= \sum_{i=0}^n \sum_{j=0}^m
    a_ib_j \sum_{k=0}^j \binom{j}{k} z^{j-k} \frac{1}{i+k+1} \left(u^{i+k+1} -
      l^{i+k+1}\right) \\ &= \sum_{i=0}^n \sum_{j=0}^m\, \sum_{k=0}^j
    \binom{j}{k} \frac{a_ib_j}{i+k+1} \left(u^{i+k+1} - l^{i+k+1}\right)
    z^{j-k}.
  \end{split}
\end{equation}

\section{Estimation problems involving polynomial distributions}\label{sc:4}

Consider the problem of estimating coefficients, $\va$, of the polynomial PDF,
$p_n(x;\va)$. For $M$ independent measurements, $x_m$, the likelihood function
is,
\begin{equation}\label{eq:130}
  L(\vx;\va) = \prod_{m=1}^M p_n(x_m;\va) = \prod_{m=1}^M \sum_{i=0}^n a_i
  x_m^i = \prod_{m=1}^M \va^T \vx_m
\end{equation}
where the column vector, $\vx_m=[x_m^i]$, $i=0,1,\ldots,n$. The maximum
likelihood (ML) estimation sets the first derivative of the log-likelihood
equal to zero, i.e.,
\begin{equation}
  \frac{\partial}{\partial \va} L(\vx;\va) \overset{!}{=} \Zs \quad
  \Leftrightarrow \quad \frac{\partial}{\partial \va} \sum_{m=1}^M \log(\va^T
  \vx_m) = \sum_{m=1}^M \frac{\vx_m^T}{\va^T \vx_m} \overset{!}{=} \Zs. 
\end{equation}

Another strategy to maximize the likelihood \eref{eq:130} is by assuming the
cosine theorem, i.e.,
\begin{equation}
  \argmax_{\va} \prod_{m=1}^M \va^T \vx_m  = \argmax_{\va} \prod_{m=1}^M
  \frac{\va^T}{\norm{\va}} \frac{\vx_m}{\norm{\vx_m}} =
  \argmax_{\va} \prod_{m=1}^M \cos \phi_m.
\end{equation}
Using the geometric-arithmetic mean inequality \cite{Aldaz2009}, the likelihood
is maximized when the vector of coefficients, $\va$, is aligned with all the
observations, $\vx_m$. This can be approximated by assuming that the distances
between the normalized vectors, $\va^T/\norm{\va}$, and, $\vx_m/\norm{\vx_m}$,
are constant, for $\forall m$. Then, the estimate is the centroid of
observations, i.e., $\hat{\va}= \frac{1}{M} \sum_{m=1}^M \vx_m/\norm{\vx_m}$.

For $M=2$, or, equivalently, only two out of $M$ measurements are considered at
a time, the ML estimator can be constrained as in \eref{eq:110}, i.e.,
\begin{equation}
  \hat{\va}_{12}= \argmax_{\va} \va^T \vX_{12} \va, \quad \mbox{s.t.}\quad
  \vw^T \va=1
\end{equation}
where $\vx_i=[ x_i]$, $j=0,1,\ldots,n$, and $\vX_{12}= \vx_1 \vx_2^T$, is a
$(n+1)\times(n+1)$ square matrix. The first derivative of the corresponding
Lagrangian must be equal to zero, i.e.,
\begin{equation}
  \begin{split}
    & \frac{\partial}{\partial \va} \left(  \va^T \vX_{12} \va + \lambda (\vw^T
      \va -1) \right) \overset{!}{=} \Zs \\
    & \Rightarrow \quad \va = -\frac{\lambda}{2} \vX_{12}^{-1} \vw,\quad
    \lambda = \frac{-2}{\vw^T \vX_{12}^{-1} \vw}.
  \end{split}
\end{equation}
Consequently, the ML estimate is,
\begin{equation}
  \hat{\va}_{12}= \frac{\vX_{12}^{-1}\vw}{\vw^T \vX_{12}^{-1}\vw}
\end{equation}
and its likelihood is equal to, $\left(\vw^T \vX_{12}^{-1}\vw\right)^{-1}$.
Finally, the observation pairs, $(\vx_1,\vx_2)$, $(\vx_3,\vx_4)$, $\ldots$, are
independent, so the final ML estimate is,
\begin{equation}
  \hat{\va} = \frac{2}{M}\sum_{i=1}^{M/2} \frac{\vX_{2i-1,2i}^{-1}\vw}{\vw^T
    \vX_{2i-1,2i}^{-1}\vw}.
\end{equation}

The Cramer-Rao bound lower-bounds the covariance matrix of the estimation
error, $\hat{\va}-\va$, of any unbiased estimator, i.e.,
\begin{equation}
  \cov{\hat{\va}-\va} \overset{\E{\hat{\va}}=\va}{=}
  \var{\hat{\va}} \geq \vJ^{-1}(\va)
\end{equation}
where $\vJ(\va)$ is the Fisher information matrix. In order to calculate the
the elements of this matrix, it is useful to instead assume Form II of
polynomial distribution, $p_n(x;\vr)= a_n\prod_{i=1}^n (x-r_i)$, and the
problem of estimating the parameters, $\vr$, i.e.,
\begin{equation}\label{eq:120}
  \begin{split}
    [\vJ]_{ij} &= \E{ \left(\frac{\partial}{\partial r_i} \log p_n(x;\vr)
      \right) \left(\frac{\partial}{\partial r_j} \log p_n(x;\vr) \right) } \\
    &= \E{ \left(\frac{\partial}{\partial r_i} \log (x-r_i) \right)
      \left(\frac{\partial}{\partial r_j} \log (x-r_j) \right) } \\
    &= \E{ \frac{1}{(x-r_i)} \frac{1}{(x-r_j)} } = \int_l^u a_n
    \prod_{\substack{k=1\\ k\neq i,j}}^n (x-r_k) \df x. 
  \end{split}
\end{equation}
The last integral in \eref{eq:120} can be computed by converting the Form II
polynomial into Form I.

The coefficients, $\va$, can be also estimated by the method of moments
\cite{Munkhammar2017, Mnatsakanov2009}. In particular, the $k$-th general
moment of a polynomial distribution, $p_n(x)$, is, (cf. \apref{Apx:A})
\begin{equation}
  \cM_k = \int_l^u x^k \sum_{i=0}^n a_i x^i \df x = \sum_{i=0}^n
  \frac{a_i}{i+k+1} \left( u^{i+k+1} - l^{i+k+1} \right).
\end{equation}
Observing a vector of the first $K$ general moments, $\vM=[\cM_k]$,
$k=1,2,\ldots,K$, and pre-computing the matrix,
$\vB=[(i+k+1)^{-1} (u^{i+k+1} - l^{i+k+1})]$, $i=0,1,\ldots,n$, the estimation
can be again defined as constrained or unconstrained least-square regression,
i.e.,
\begin{equation}
  \hat{\va} = \argmin_{\va} \norm{\vM - \vB \va},\quad \mbox{s.t.}\quad
  \vw^T\va = 1
\end{equation}
which can be efficiently solved as in \eref{eq:140}.

Finally, Bayesian estimation methods for estimating the coefficients, $\va$,
require adopting a prior, $p(\va)$. Since the coefficients are likely to be
mutually correlated, defining such prior distribution may be challenging,
unless a Gaussian prior can be assumed. On the other hand, consider a general
probabilistic model with observations $X$ and a parameter, $\theta$, which is
described by the likelihood, $p(X|\theta)$, and the prior, $p(\theta)$. If the
likelihood and the prior are both polynomially distributed, then the
corresponding posterior, $p(\theta|X)\propto p(X|\theta)p(\theta)$, is also
polynomially distributed.

\section{Generating polynomially distributed random variables}\label{sc:5}

Like most other distributions, polynomial distributions cannot be easily
inverted. Then, it is challenging to use the inverse method for generating
random variables. On the other hand, the CDF of a polynomially distributed
random variable is another polynomial as shown in \apref{Apx:A}. A CDF
discretization can be then used as a general strategy for implementing the
inverse method of generating random variables from a distribution with a known
CDF. In particular, let approximate the CDF by a piecewise linear function
between the samples, $(x_i,F(x_i))$, $i=1,2,\ldots$. The inverse value,
$X=F^{-1}(U)$, where $U\in(0,1)$ is a uniformly distributed random variable, is
then approximated as,
\begin{equation}
  x = x_i + \frac{(x_{i+1}-x_i)(u-F(x_i))}{F(x_{i+1}-F(x_i))}.
\end{equation}

Discretization can be also assumed when constructing a proposal distribution
for the rejection sampling method. In particular, the proposal can be either a
piecewise linear or piecewise step-wise function defined by the samples,
$(x_i,f(x_i))$, $i=1,2,\ldots$. This allows defining the proposal distribution
closely matching the target distribution, which can considerably increase the
sampling efficiency.

\section{Conclusion}\label{sc:6}

Polynomials are often used for approximating univariate and multivariate
functions including probability distributions. This paper defined polynomial
distributions, which can be also used to approximate other canonical and
empirically estimated distributions over finite intervals of support.
Polynomial distributions can be considered to be a more flexible alternative to
commonly used canonical distributions. More importantly, in this paper, many
key properties of polynomial distributions were derived and presented as
closed-form expressions.

There is a need for defining family of distributions such as polynomial
distributions that can be used more universally for solving problems in
probability, statistics and data analysis. The polynomial distributions
considered in this paper are univariate and continuous; the extension to
multivariate and discrete polynomial distributions may be subject of our future
work. Polynomials could be generalized as a weighted linear sum of non-linear
functions of the same variable. A number of research problems remain open. For
example, given a polynomial, identify all sub-intervals where it is
non-negative. Or, given a polynomial order and an interval of support, the task
is to find all polynomials that represent a PDF. This problem can be further
constrained by the desired number of modes, the smoothness and/or sparsity
conditions, and assuming other statistical and algebraic properties of the
polynomial distributions. Moreover, the problem of determining the minimum
polynomial degree or sparsity to satisfy given constraints has not been
considered in this paper. It would be also very useful to investigate how to
interpret polynomial distributions, especially as they may arise naturally when
observing some stochastic phenomena.

\vspace{6pt} 

\funding{This research was funded by a research grant from Zhejiang University.}

\clearpage
\appendixtitles{yes}
\appendix

\newpage\section{Basic properties of Form I polynomials}\label{Apx:A}\unskip

\apxitem{Definition}
\begin{equation}
  p_n(x)=\sum_{i=0}^n a_i\,x^i, \quad a_i\in\R,\ a_n\neq 0
\end{equation}

\apxitem{Roots, $n=1$}
\begin{equation}
  p_1(x)=0 \quad\Leftrightarrow\quad x_1 = -\frac{a_0}{a_1}
\end{equation}

\apxitem{Roots, $n=2$}
\begin{equation}
  p_2(x)=0  \quad\Leftrightarrow\quad \begin{array}{lc}
    x_{1,2}=\frac{-a_1\pm \sqrt{a_1^2-4a_2 a_0}}{2 a_2} & a_1^2 > 4a_2 a_0 \\
    x_1=x_2=-\frac{a_1}{2 a_2} & a_1^2 = 4a_2 a_0 \\ x_{1,2} \notin \R & a_1^2
    < 4a_2 a_0  \end{array}
\end{equation}

\apxitem{Roots, $n=3$}
\begin{equation}
  p_3(x)=0  \quad\Leftrightarrow\quad \begin{array}{lc}
    x_1 =  \sqrt[3]{-\frac{b}{2}+\sqrt{\frac{b^2}{4}+\frac{a^3}{27}}} +
    \sqrt[3]{-\frac{b}{2}-\sqrt{\frac{b^2}{4}+\frac{a^3}{27}}},\
    x_{2,3}\notin\R & D<0 \\
    x_1=x_2=x_3 = -\frac{a_2}{3a_3} & D=0,\ a_2^2=3a_3 a_1 \\
    x_1=x_2= \frac{9a_3 a_0-a_2 a_1}{2(a_2^2-3a_3 a_1)},\ x_3=
    \frac{4a_3 a_2 a_0-9a_3^2 a_0-a_2^3}{a_3(a_2^2-3a_3 a_1)} & D=0,\
    a_2^2\neq 3a_3 a_1 \\
    x_k = -\frac{1}{3a}(b+\xi^{k-1}C+\frac{\Delta_0}{\xi^{k-1}C}),\ k=1,2,3
    & D>0     
  \end{array}
\end{equation}
where
\begin{equation*}
  a = \frac{-a_2}{3a_3},\quad b = a^3+\frac{a_2 a_1-3a_3 a_0}{6a_3^2},\quad
  D = \frac{4(a_2^2-3a_3 a_1)-(2a_2^3-9a_3 a_2 a_1+27a_3^2 a_0^2)}{27a_3^2}
\end{equation*}
\begin{equation*}
  \Delta_0 = a_2^2-3a_3,\quad \Delta_1 = 2a_2^3-9a_3 a_2 a_1+27a_3^2a_0,\quad
  C = \sqrt[3]{\frac{\Delta_1\pm\sqrt{\Delta_1^2-4\Delta_0^3}}{2}},\quad
  \xi = \frac{-1+\sqrt{-3}}{2}
\end{equation*}

\apxitem{Roots, general case}

\begin{itemize}
\item By Abel–Ruffini's theorem, closed-form expressions for roots of a
  polynomial exist for polynomials of degree at most $n=4$, and there is no
  algebraic solution for the polynomial roots for degree $n>4$.
\item The total number of real roots of a polynomial within a given interval or
  over all real numbers can be determined by Sturm's theorem.
\item Other relationships between polynomial coefficients and roots can be
  obtained such as the Vieta's formulas:
  \begin{equation}
    \sum_{i_1\neq i_2 \neq \cdots \neq i_k} r_{i_1}r_{i_2}\cdots r_{i_k} = (-1)^i\,
    \frac{a_{n-i}}{a_n},\quad k\leq i=1,2,\ldots,n
  \end{equation}
\end{itemize}

\apxitem{Indefinite integral}
\begin{equation}
  \Ip_n(x) \equiv \int p_n(x) \df x = \sum_{i=0}^n \ \frac{a_i}{i+1} x^{i+1}
\end{equation}

\apxitem{Definite integral}
\begin{equation}
  P_n(u) \equiv \int_{-\infty}^u p_n(x) \df x  
\end{equation}
\begin{equation}
  \begin{split}
    \int_{l}^{u} p_n(x) \df x &= \int_{-\infty}^u p_n(x)\df x -
    \int_{-\infty}^l p_n(x) \df x = P_n(u) - P_n(l),\quad l<u \\
    &= \sum_{i=0}^n \ \frac{a_i}{i+1} \left(u^{i+1}- l^{i+1}\right)
    \equiv \Ip(u) - \Ip(l)
  \end{split}
\end{equation}

\apxitem{Indefinite $k$-fold integral, $k>1$}
\begin{equation}
  \idotsint\limits_k p_n(x) \df x^k = \sum_{i=0}^n
  \frac{a_i}{(i+1)(i+2)\dots(i+k)}\, x^{i+k} 
  = \sum_{i=0}^n a_i\,\frac{i!}{(i+k)!}\, x^{i+k}
\end{equation}

\apxitem{Derivative}
\begin{equation}
  \Dp_n(x) = \frac{\df}{\df x} p_n(x) = \sum_{i=1}^n i \,a_i \, x^{i-1}
\end{equation}

\apxitem{$k$-th derivative, $1< k \leq n$}
\begin{equation}
\begin{split}
  p^{(k)}_n(x) & = \sum_{i=k}^n i(i-1)\dots (i-k+1) \, a_i
  \, x^{i-k} \\ &= \sum_{i=k}^n a_i \, \frac{i!}{(i-k)!} \, x^{i-k}
  =  \sum_{i=k}^n a_i \, k! \binom{i}{k} \, x^{i-k}
\end{split}
\end{equation}

\apxitem{$k$-th moment, $k\geq 1$}
\begin{equation}
  \begin{split}
    \int x^k\,p_n(x)\df x &= \sum_{i=0}^n \ \frac{a_i}{i+k+1} x^{i+k+1} \\
    \int_l^u x^k\,p_n(x)\df x &= \sum_{i=0}^n \ \frac{a_i}{i+k+1} \left(
      u^{i+k+1} - l^{i+k+1} \right)
  \end{split}
\end{equation}

\apxitem{Characteristic function}
\begin{equation}
  \begin{split}
    \phi_X(t) &= \E{\eee^{\jj t X}} = \int_{-1}^{+1} p_n(x) \eee^{\jj t x} \df
    x = \sum_{i=0}^n a_i \int_{-1}^{+1} x^i \eee^{\jj t x} \df x \\
    &= \sum_{i=0}^n a_i\,\frac{\jj}{t} \int_{-t/\jj}^{t/\jj}
    \left(\frac{x\jj}{t}\right)^i \eee^{-x} \df x \\
    &= \sum_{i=0}^n a_i \left(\frac{\jj}{t}\right)^{i+1}
    \left( \int_{-t/\jj}^{\infty} x^i \eee^{-x} \df x -
      \int_{t/\jj}^{\infty} x^i \eee^{-x} \df x \right) \\ 
    &= \sum_{i=0}^n \frac{a_i}{(-\jj t)^{i+1}} \left( \Gamma(1+i,\jj t) -
      \Gamma(1+i,-\jj t) \right)
  \end{split}
\end{equation}
where $\Gamma(a,z)$ is the incomplete Gamma function 

\newpage\section{Basic properties of Form II polynomials}\label{Apx:B}\unskip

\apxitem{Definition}
\begin{equation}
  p_n(x) = a_n \prod \limits_{i=1}^n (x-r_i),\quad a_n\neq 0,\ r_i\in\Cb
\end{equation}

\apxitem{Recursive form}
\begin{equation}
  \begin{split} p_n(x) &= \frac{a_n}{a_{n-1}}\, (x-r_n)\, p_{n-1}(x),\quad n>1
    \\ p_1(x) &= x-r_1
  \end{split}
\end{equation}

\apxitem{Converting Form I to Form II, general case}
\begin{equation}
  \sum_{i=0}^n a_i x^i=a_n \prod \limits_{i=1}^n (x-r_i)
\end{equation}
\begin{equation}
  \begin{split}
    a_n &\equiv a_n,\quad
    a_{n-1} = a_n {(-1)}^1 \sum_{i=1}^N r_i,\quad
    a_{n-2} = a_n {(-1)}^{2} \sum_{i=1,j=1,i\neq j}^N r_ir_j \\
    a_{n-3} &= a_n {(-1)}^{3} \sum_{i=1,j=1,k=1,i\neq j\neq k}^N r_ir_jr_k ,
    \quad \cdots \quad, a_0 = a_n {(-1)}^n \prod_{i = 1}^{n}r_i
  \end{split}
\end{equation}

\apxitem{Converting Form I to Form II, $n=2$}
\begin{equation}
  a_2 \equiv a_2,\quad a_1 = -a_2(r_1+r_2),\quad a_0 = a_2 r_1 r_2
\end{equation}

\apxitem{Converting Form I to Form II, $n=3$}
\begin{equation}
  a_3 \equiv a_3,\quad a_2 = -a_3(r_1+r_2+r_3),\quad 
  a_1 = a_3 (r_1 r_2+r_1 r_3+r_2 r_3),\quad a_0 = -a_3 r_1 r_2 r_3
\end{equation}

\apxitem{Converting Form I to Form II, $n=4$}
\begin{equation}
  \begin{split}
    a_4 \equiv a_4,\quad
    a_3 = & -a_4(r_1+r_2+r_3+r_4),\quad 
    a_2 = a_4(r_1r_2+r_1r_3+r_1r_4+r_2r_3+r_2r_4+r_3r_4)\\
    a_1 = & -a_4(r_1r_2r_3+r_1r_2r_4+r_1r_3r_4+r_2r_3r_4),\quad 
    a_0 = a_4 r_1 r_2 r_3 r_4
  \end{split}
\end{equation}

\apxitem{Indefinite integral} (recursive form)
\begin{equation}
  \begin{split}
    I_n(x) &= \frac{1}{a_n} \int p_n(x) \df x, \quad n>1 \\
    &= (x-r_n)I_{n-1}(x)-\int I_{n-1}(x) \df x \\
    I_1(x)&=\frac{1}{a_1}\int p_1(x) \df x = \int (x-r_1) \df x =
    \frac{1}{2} x^2 - r_1 x
  \end{split}
\end{equation}

\apxitem{Indefinite $k$-fold integral}
\begin{equation}
  \begin{split}
    I_n(x) &= (x-r_n)I_{n-1}(x)-\int I_{n-1}(x) \df x \\
    \int I_n(x) \df x &= \int (x-r_n)I_{n-1}(x) \df x - \iint I_{n-1}(x) \df
    x^2 \\ &= (x-r_n) \int I_{n-1}(x)\df x - 2 \iint I_{n-1}(x)\df x^2  \\
    \idotsint\limits_k I_n(x)\df x^k & = (x-r_n)\idotsint\limits_k I_{n-1}(x)
    \df x^k - (k+1) \idotsint\limits_{k+1} I_{n-1}(x) \df x^{k+1}
  \end{split}
\end{equation}

\apxitem{Derivative}
\begin{equation}
  \begin{split}
    \Dp_n(x) &= a_n \prod \limits_{i=1}^{n-1} (x-r_i)+ \frac{a_n}{a_{n-1}}
    (x-r_n) \Dp_{n-1}(x),\quad n>1 \\ &= \frac{a_n}{a_{n-1}} p_{n-1}(x) +
    \frac{a_n}{a_{n-1}} (x-r_n) \Dp_{n-1}(x) \\
    \Dp_1(x) &= a_1
  \end{split}
\end{equation}

\apxitem{$k$-th derivative, $1< k \leq n$}
\begin{equation}
  \begin{split}
    \Dp_n(x) &= \frac{a_n}{a_{n-1}}\, p_{n-1}(x) +
    \frac{a_n}{a_{n-1}}(x-r_n)\, \Dp_{n-1}(x) \\
    p_n^{(k)}(x) &= k \frac{a_n}{a_{n-1}}\, p_{n-1}^{(k-1)}(x) +
    \frac{a_n}{a_{n-1}}(x-r_n)\, p_{n-1}^{(k)}(x)
  \end{split}
\end{equation}

\apxitem{$k$-th moment, $k\geq 1$}
\begin{equation}
  \begin{split}
    \Intn{}{p}_n(x) \equiv \int p_n(x) \df x &,\quad 
    \Intn{k}{p}_n(x) \equiv \idotsint\limits_k p_n(x) \df x \\    
    \int x^k\,p_n(x)\df x &= x^k\, \Intn{}{p}_{n}(x) - k \int x^{k-1}
    \Intn{}{p}_n(x)\df x^2 \\ \int x^{k-1}\, \Intn{}{p}_n(x) \df x &=
    x^{k-1} \Intn{2}{p}_n(x) - (k-1) \int x^{k-2}\, \Intn{2}{p}_n(x) \df x \\
    & \ \vdots \\ \int x \Intn{k-1}{p}_n(x)\df x &= x \Intn{k}{p}_n(x) -
    \Intn{k+1}{p}_n(x) \\ 
    \int_l^u x^k\,p_n(x)\df x =& \left[ x^k \Intn{}{p}_{n}(x) \right]_l^u
    - k \int_l^u x^{k-1} \Intn{}{p}_n(x)\df x
  \end{split}
\end{equation}

\apxitem{Moment generating function}
\begin{equation}
  \begin{split}
    M(t) = \int_l^u e^{tx} p_n(x) \df x &= \frac{e^{tx}}{t} p_n(x)- \frac{1}{t}
    \int_l^u e^{tx} \Dern{1}{p}_n(x) \df x \\
    \int_l^u e^{tx} \Dern{1}{p}_n(x) \df x &= \frac{e^{tx}}{t} p'_n(x) -
    \frac{1}{t} \int_l^u e^{tx} \Dern{2}{p}_n(x) \df x \\
    & \ \vdots \\
    \int_l^u e^{tx}\Dern{n}{p}_n(x) \df x &= \int_l^u e^{tx} a_n n! \df x
    =\frac{a_n n!}{t} e^{t^u-t^l}
  \end{split}
\end{equation}

\newpage\section{Basic properties of Form III polynomials}\label{Apx:C}\unskip

\apxitem{Definition}
\begin{equation}
  p_n(x) = \frac{s_m(x)}{q_n(x)} = \frac{s_m(x)}{c_n \prod_{i=1}^n
    (x-r_i)} = \sum_{i=1}^n \frac{a_i}{x-r_i},\quad m<n,\ c_n\neq 0,\ r_i\neq
  r_j\ \forall i\neq j
\end{equation}
where the residuals, $a_i = \frac{s_m(r_i)}{\dot{q}_n(r_i)} \neq 0$

\apxitem{Indefinite integral}
\begin{equation}
  \begin{split}
    r_i\in\R:\ \int_l^u \frac{1}{x-r_i} \df x &= \left\{ \begin{array}{cc}
        \ln \frac{u-r_i}{l-r_i}, & r_i<l \mbox{ or } r_i> u \\
        n.c. & \mbox{otherwise} \\ \end{array} \right. \\
    r_i\in\Cb: \int_l^u \frac{1}{x-r_i} \df x &=
    \left\{ \begin{array}{cc} \ln \frac{u-r_i}{l-r_i}, & \Re{r_i}<l\mbox{ or }
        \Re{r_i}> u \mbox{ or } \Im{r_i}\neq 0 \\ n.c. & \mbox{otherwise}
        \\ \end{array} \right. \\
    \int \sum_{i=1}^n \frac{c_i}{x-r_i} \df x &= \sum_{i=1}^n c_i \ln (x-r_i)
  \end{split}
\end{equation}

\apxitem{Definite integral}
\begin{equation}
  \int_l^u \sum_{i=1}^n \frac{c_i}{x-r_i} \df x = \sum_{i=1}^n c_i\,
  \frac{\ln (u-r_i)}{\ln (l-r_i)}
\end{equation}

\apxitem{$k$-th derivative, $1\leq k\leq n$}
\begin{equation}
  p^{(k)}_n(x) = (-1)^k \,k! \sum_{i=1}^n\frac{c_i}{(x-r_i)^{k+1}}
\end{equation}

\apxitem{$k$-fold integral, $k\geq 1$}
\begin{equation}
  \frac{\df^k}{\df x^k} \left(\frac{x^{k-1}\ln x}{(k-1)!}\right) =\frac{1}{x}
\end{equation}
\begin{equation}
  \idotsint\limits_k p_n(x) \df x^k= \sum_{i=1}^n c_i \,\frac{(x-r_i)^{k-1}
    \ln(x-r_i)}{(k-1)!}
\end{equation}

\apxitem{$k$-th moment, $k\geq 1$}
\begin{equation}
  \begin{split}
    \int_{l}^{u} x^k \sum_{i=1}^n \frac{c_i}{x-r_i} \df x &=
    \sum_{i=1}^n c_i\, r_i^k\, \left( \beta_{l/r_i}(1+k,0) -
      \beta_{u/r_i}(1+k,0) \right),\ 0\leq l < u\leq 1 \\
    \int_l^u \frac{x^k}{x-r_i} \df x &=
    - r_i^{k-1} \int_l^u \left(\frac{x}{r_i}\right)^k
    \left(1-\frac{x}{r_i}\right)^{-1} \df x = - r_i^k \int_{l/r_i}^{u/r_i} x^k
    (1-x)^{-1} \df x \\ &= r_i^k\, \left( \beta_{l/r_i} (1+k,0) -
      \beta_{u/r_i} (1+k,0) \right),\ r_i\neq 0 \\
    \int_l^u \frac{x^k}{x} \df x &= \frac{1}{k} (u^k-l^k),\ r_i=0
  \end{split}
\end{equation}
where $\beta_z(a,b) = \int_0^z t^{a-1} (1-t)^{b-1}\df t$ is incomplete
$\beta$-function

\apxitem{Characteristic function}
\begin{equation}
    \phi_X(t) = \E{\eee^{\jj t X}} = \sum_{i=1}^n c_i \int_{l}^{u}
    \frac{\eee^{\jj t x}}{x-r_i} \df x 
    = \sum_{i=1}^n c_i \eee^{\jj r_i t} \left(\Gamma(0,\jj t (r_i-l)) -
      \Gamma(0,\jj t (r_i-u)) \right),\ r\in (l,u) 
\end{equation}

\reftitle{References}
\externalbibliography{yes}
\bibliography{refer}

\begin{thebibliography}{-------}
\providecommand{\natexlab}[1]{#1}

\bibitem[Phillips and Taylor(1996)]{Phillips1996}
Phillips, G.M.; Taylor, P.J., "Best" Approximation.
\newblock In {\em Theory and Applications of Numerical Analysis}; Academic
  Press, London, UK,  1996; pp. 86--130.
\newblock
  doi:{\changeurlcolor{black}\href{https://doi.org/10.1016/b978-012553560-1/50006-9}{\detokenize{10.1016/b978-012553560-1/50006-9}}}.

\bibitem[Cheney(1982)]{Cheney1982}
Cheney, E.W.
\newblock {\em Introduction to Approximation Theory}, 2nd ed.; AMS Chelsea
  Publishing, Providence, Rhode Island,  1982.

\bibitem[Freedman and Diaconis(1981)]{Freedman1981}
Freedman, D.; Diaconis, P.
\newblock On the histogram as a density estimator:L2 theory.
\newblock {\em Zeitschrift f\"ur Wahrscheinlichkeitstheorie und Verwandte
  Gebiete} {\bf 1981}, {\em 57},~453--476.
\newblock
  doi:{\changeurlcolor{black}\href{https://doi.org/10.1007/BF01025868}{\detokenize{10.1007/BF01025868}}}.

\bibitem[Munkhammar \em{et~al.}(2017)Munkhammar, Mattsson, and
  Ryd\'en]{Munkhammar2017}
Munkhammar, J.; Mattsson, L.; Ryd\'en, J.
\newblock Polynomial probability distribution estimation using the method of
  moments.
\newblock {\em PLOS ONE} {\bf 2017}, {\em 12},~1--14.
\newblock
  doi:{\changeurlcolor{black}\href{https://doi.org/10.1371/journal.pone.0174573}{\detokenize{10.1371/journal.pone.0174573}}}.

\bibitem[Badinelli(1996)]{Badinelli1996}
Badinelli, R.D.
\newblock Approximating probability density functions and their convolutions
  using orthogonal polynomials.
\newblock {\em European Journal of Operational Research} {\bf 1996}, {\em
  95},~211--230.
\newblock
  doi:{\changeurlcolor{black}\href{https://doi.org/10.1016/0377-2217(95)00250-2}{\detokenize{10.1016/0377-2217(95)00250-2}}}.

\bibitem[Abdous and Bensaid(2007)]{Abdous2007}
Abdous, B.; Bensaid, E.
\newblock Multivariate local polynomial fitting for a probability distribution
  function and its partial derivatives.
\newblock {\em Journal of Nonparametric Statistics} {\bf 2007}, {\em
  13},~77--94.
\newblock
  doi:{\changeurlcolor{black}\href{https://doi.org/10.1080/10485250008832843}{\detokenize{10.1080/10485250008832843}}}.

\bibitem[Gasca and Sauer(2000)]{Gasca2000}
Gasca, M.; Sauer, T.
\newblock Polynomial interpolation in several variables.
\newblock {\em Advances in Computational Mathematics} {\bf 2000}, {\em
  12},~377--410.
\newblock
  doi:{\changeurlcolor{black}\href{https://doi.org/10.1023/A:1018981505752}{\detokenize{10.1023/A:1018981505752}}}.

\bibitem[Ghasemi and Marshall(2010)]{Ghasemi2010}
Ghasemi, M.; Marshall, M.
\newblock Lower bounds for a polynomial in terms of its coefficients.
\newblock {\em Archiv der Mathematik} {\bf 2010}, {\em 95},~343--353.
\newblock
  doi:{\changeurlcolor{black}\href{https://doi.org/10.1007/s00013-010-0179-0}{\detokenize{10.1007/s00013-010-0179-0}}}.

\bibitem[Forsythe(1957)]{Forsythe1957}
Forsythe, G.E.
\newblock Generation and Use of Orthogonal Polynomials for Data-Fitting with a
  Digital Computer.
\newblock {\em Journal of the Society for Industrial and Applied Mathematics}
  {\bf 1957}, {\em 5},~74--88.
\newblock
  doi:{\changeurlcolor{black}\href{https://doi.org/10.1137/0105007}{\detokenize{10.1137/0105007}}}.

\bibitem[Cunis(2018)]{Cunis2018}
Cunis, T.
\newblock The pwpfit Toolbox for Polynomial and Piece-wise Polynomial Data
  Fitting.
\newblock  In Proc. International Federation of Automatic Control,  2018, pp.
  682--687.
\newblock
  doi:{\changeurlcolor{black}\href{https://doi.org/10.1016/j.ifacol.2018.09.204}{\detokenize{10.1016/j.ifacol.2018.09.204}}}.

\bibitem[Hiang and Ali(2013)]{Hiang2013}
Hiang, T.S.; Ali, J.M.
\newblock Quartic and quintic polynomial interpolation.
\newblock  In AIP Conference Proceedings,  2013, Vol. 1522, pp. 664--675.
\newblock
  doi:{\changeurlcolor{black}\href{https://doi.org/10.1063/1.4801189}{\detokenize{10.1063/1.4801189}}}.

\bibitem[Gao \em{et~al.}(2020)Gao, Ji, Zhang, Shao, Wang, and Shi]{Gao2020}
Gao, J.; Ji, W.; Zhang, L.; Shao, S.; Wang, Y.; Shi, F.
\newblock Fast Piecewise Polynomial Fitting of Time-Series Data for Streaming
  Computing.
\newblock {\em IEEE Access} {\bf 2020}, {\em 8},~43764--43775.
\newblock
  doi:{\changeurlcolor{black}\href{https://doi.org/10.1109/ACCESS.2020.2976494}{\detokenize{10.1109/ACCESS.2020.2976494}}}.

\bibitem[Guo \em{et~al.}(2020)Guo, Narayan, and Zhou]{Guo2020}
Guo, L.; Narayan, A.; Zhou, T.
\newblock Constructing Least-Squares Polynomial Approximations.
\newblock {\em SIAM Review} {\bf 2020}, {\em 62},~483--508.
\newblock
  doi:{\changeurlcolor{black}\href{https://doi.org/10.1137/18M1234151}{\detokenize{10.1137/18M1234151}}}.

\bibitem[Han \em{et~al.}(2007)Han, Liu, and Ji]{Han2007}
Han, H.; Liu, H.; Ji, X.
\newblock Interpolation to Data Points in Plane with Cubic Polynomial
  Precision.
\newblock  In Proc. Technologies for E-Learning and Digital Entertainment,
  2007, pp. 677--686.
\newblock
  doi:{\changeurlcolor{black}\href{https://doi.org/10.1007/978-3-540-73011-8_65}{\detokenize{10.1007/978-3-540-73011-8_65}}}.

\bibitem[Melucci(2019)]{Melucci2019}
Melucci, M.
\newblock A brief survey on probability distribution approximation.
\newblock {\em Computer Science Review} {\bf 2019}, {\em 33},~91--97.
\newblock
  doi:{\changeurlcolor{black}\href{https://doi.org/10.1016/j.cosrev.2019.06.001}{\detokenize{10.1016/j.cosrev.2019.06.001}}}.

\bibitem[He \em{et~al.}(2021)He, Hao, and Li]{He2021}
He, W.; Hao, P.; Li, G.
\newblock A novel approach for reliability analysis with correlated variables
  based on the concepts of entropy and polynomial chaos expansion.
\newblock {\em Mechanical Systems and Signal Processing} {\bf 2021}, {\em
  146},~106980.
\newblock
  doi:{\changeurlcolor{black}\href{https://doi.org/10.1016/j.ymssp.2020.106980}{\detokenize{10.1016/j.ymssp.2020.106980}}}.

\bibitem[Chen \em{et~al.}(2015)Chen, Yuan, Hua, Zheng, and Wang]{Chen2015}
Chen, D.; Yuan, Z.; Hua, G.; Zheng, N.; Wang, J.
\newblock Similarity Learning on an Explicit Polynomial Kernel Feature Map for
  Person Re-Identification.
\newblock  In Proc. IEEE Conference on Computer Vision and Pattern Recognition,
   2015.
\newblock
  doi:{\changeurlcolor{black}\href{https://doi.org/10.1109/CVPR.2015.7298764}{\detokenize{10.1109/CVPR.2015.7298764}}}.

\bibitem[Cotter(1990)]{Cotter1990}
Cotter, N.E.
\newblock The Stone-Weierstrass theorem and its application to neural networks.
\newblock {\em IEEE Transactions on Neural Networks} {\bf 1990}, {\em
  1},~290--295.
\newblock
  doi:{\changeurlcolor{black}\href{https://doi.org/10.1109/72.80265}{\detokenize{10.1109/72.80265}}}.

\bibitem[Tong \em{et~al.}(2021)Tong, Yu, Li, Liu, Qin, and Li]{Tong2021}
Tong, Y.; Yu, L.; Li, S.; Liu, J.; Qin, H.; Li, W.
\newblock Polynomial Fitting Algorithm Based on Neural Network.
\newblock {\em ASP Transactions on Pattern Recognition and Intelligent Systems}
  {\bf 2021}, {\em 1},~32--39.
\newblock
  doi:{\changeurlcolor{black}\href{https://doi.org/10.52810/TPRIS.2021.100019}{\detokenize{10.52810/TPRIS.2021.100019}}}.

\bibitem[Barbeau(1989)]{Barbeau1989}
Barbeau, E.J.
\newblock {\em Polynomials}; Springer-Verlag, New York, USA,  1989.

\bibitem[Rahman and Schmeisser(2002)]{Rahman2002}
Rahman, Q.I.; Schmeisser, G.
\newblock {\em Analytic Theory of Polynomials}; Oxford University Press, New
  York, USA,  2002.

\bibitem[Apostol(1974)]{Apostol1974}
Apostol, T.M.
\newblock {\em Mathematical Analysis}, 2nd ed.; Addison-Wesley, Reading, MA,
  USA,  1974.

\bibitem[Rahman(2009)]{Rahman2009}
Rahman, S.
\newblock An extended polynomial dimensional decomposition method for arbitrary
  probability distributions.
\newblock {\em Journal of Engineering Mechanics} {\bf 2009}, {\em
  135},~1439--1451.
\newblock
  doi:{\changeurlcolor{black}\href{https://doi.org/10.1061/(ASCE)EM.1943-7889.0000047}{\detokenize{10.1061/(ASCE)EM.1943-7889.0000047}}}.

\bibitem[Funaro(1992)]{Funaro1992}
Funaro, D.
\newblock {\em Polynomial Approximation of Differential Equations}; Springer,
  Berlin, Germany,  1992.

\bibitem[Guo \em{et~al.}(2009)Guo, Shen, and Wang]{Guo2009}
Guo, B.Y.; Shen, J.; Wang, L.L.
\newblock Generalized Jacobi Polynomials/Functions and Their Applications.
\newblock {\em Applied Numerical Mathematics} {\bf 2009}, {\em 59},~1011--1028.
\newblock
  doi:{\changeurlcolor{black}\href{https://doi.org/10.1016/j.apnum.2008.04.003}{\detokenize{10.1016/j.apnum.2008.04.003}}}.

\bibitem[Boas and Klamkin(1977)]{Boas1977}
Boas, R.P.; Klamkin, M.S.
\newblock Extrema of Polynomials.
\newblock {\em Mathematics Magazine} {\bf 1977}, {\em 50},~75--78.

\bibitem[Hanzon and Jibetean(2003)]{Hanzon2003}
Hanzon, B.; Jibetean, D.
\newblock Global Minimization of a Multivariate Polynomial using Matrix
  Methods.
\newblock {\em Journal of Global Optimization} {\bf 2003}, {\em 27},~1--23.
\newblock
  doi:{\changeurlcolor{black}\href{https://doi.org/10.1023/A:1024664432540}{\detokenize{10.1023/A:1024664432540}}}.

\bibitem[Qi and Teo(2003)]{Qi2003}
Qi, L.; Teo, K.L.
\newblock Multivariate Polynomial Minimization and Its Application in Signal
  Processing.
\newblock {\em Journal of Global Optimization} {\bf 2003}, {\em 26},~419--433.
\newblock
  doi:{\changeurlcolor{black}\href{https://doi.org/10.1023/A:1024778309049}{\detokenize{10.1023/A:1024778309049}}}.

\bibitem[Uteshev and Cherkasov(1998)]{Uteshev1998}
Uteshev, A.Y.; Cherkasov, T.M.
\newblock The Search for the Maximum of a Polynomial.
\newblock {\em Journal Symbolic Computation} {\bf 1998}, {\em 25},~587--618.
\newblock
  doi:{\changeurlcolor{black}\href{https://doi.org/10.1006/jsco.1997.0190}{\detokenize{10.1006/jsco.1997.0190}}}.

\bibitem[Pan(1997)]{Pan1997}
Pan, V.Y.
\newblock Solving A Polynomial Equation: Some History And Recent Progress.
\newblock {\em SIAM Review} {\bf 1997}, {\em 39},~187--220.
\newblock
  doi:{\changeurlcolor{black}\href{https://doi.org/10.1137/S0036144595288554}{\detokenize{10.1137/S0036144595288554}}}.

\bibitem[Beji(2021)]{Beji2021}
Beji, S.
\newblock Polynomial Functions Composed of Terms with Non-Integer Powers.
\newblock {\em Advances in Pure Mathematics} {\bf 2021}, {\em 11},~791--806.
\newblock
  doi:{\changeurlcolor{black}\href{https://doi.org/10.4236/apm.2021.1110053}{\detokenize{10.4236/apm.2021.1110053}}}.

\bibitem[Papoulis and Pillai(2002)]{Papoulis02}
Papoulis, A.; Pillai, S.U.
\newblock {\em Probability, Random Variables, and Stochastic Processes}, 4th
  ed.; McGraw-Hill, New York, USA,  2002.

\bibitem[Alzaatreh \em{et~al.}(2013)Alzaatreh, Lee, and Famoye]{Alzaatreh2013}
Alzaatreh, A.; Lee, C.; Famoye, F.
\newblock A new method for generating families of continuous distributions.
\newblock {\em Metron} {\bf 2013}, {\em 71},~63--79.
\newblock
  doi:{\changeurlcolor{black}\href{https://doi.org/10.1007/s40300-013-0007-y}{\detokenize{10.1007/s40300-013-0007-y}}}.

\bibitem[Maulud and Abdulazeez(2020)]{Maulud2020}
Maulud, D.; Abdulazeez, A.M.
\newblock A Review on Linear Regression Comprehensive in Machine Learning.
\newblock {\em Journal of Applied Science and Technology Trends} {\bf 2020},
  {\em 1},~140--147.
\newblock
  doi:{\changeurlcolor{black}\href{https://doi.org/10.38094/jastt1457}{\detokenize{10.38094/jastt1457}}}.

\bibitem[Gorav \em{et~al.}(2020)Gorav, Pandey, Shukla, and
  Zisopoulos]{Gorav2020}
Gorav, J.; Pandey, A.; Shukla, H.; Zisopoulos, C.
\newblock How many zeros of a random sparse polynomial are real?
\newblock  In Proc. ISSAC,  2020, pp. 273--280.
\newblock
  doi:{\changeurlcolor{black}\href{https://doi.org/10.1145/3373207.3404031}{\detokenize{10.1145/3373207.3404031}}}.

\bibitem[Bini(1996)]{Bini1996}
Bini, D.A.
\newblock Numerical computation of polynomial zeros by means of Aberth's
  method.
\newblock {\em Numerical Algorithms} {\bf 1996}, {\em 13},~179--200.
\newblock
  doi:{\changeurlcolor{black}\href{https://doi.org/10.1007/BF02207694}{\detokenize{10.1007/BF02207694}}}.

\bibitem[Lang and Frenzel(1994)]{Lang1994}
Lang, M.; Frenzel, B.C.
\newblock Polynomial root finding.
\newblock {\em IEEE Signal Processing Letters} {\bf 1994}, {\em 1},~141--143.
\newblock
  doi:{\changeurlcolor{black}\href{https://doi.org/10.1109/97.329845}{\detokenize{10.1109/97.329845}}}.

\bibitem[Aldaz(2009)]{Aldaz2009}
Aldaz, J.M.
\newblock Self–Improvement Of The Inequality Between Arithmetic And Geometric
  Means.
\newblock {\em Journal of Mathematical Inequalities} {\bf 2009}, {\em
  3},~213--216.
\newblock
  doi:{\changeurlcolor{black}\href{https://doi.org/10.7153/jmi-03-21}{\detokenize{10.7153/jmi-03-21}}}.

\bibitem[Mnatsakanov and Hakobyan(2009)]{Mnatsakanov2009}
Mnatsakanov, R.M.; Hakobyan, A.S.
\newblock Recovery of Distributions via Moments.
\newblock  In Proc. Optimality: The Third Erich L. Lehmann Symposium,  2009,
  Vol.~57, pp. 252--265.
\newblock
  doi:{\changeurlcolor{black}\href{https://doi.org/10.1214/09-LNMS5715}{\detokenize{10.1214/09-LNMS5715}}}.

\end{thebibliography}

\end{document}